\documentclass[aps,prc,twocolumn,showpacs,superscriptaddress,floatfix,nofootinbib,10pt]{revtex4}
\usepackage{graphicx}
\usepackage{dcolumn}
\usepackage{amsfonts}
\usepackage{amsmath}
\usepackage{color}

\begin{document}

\title{Localization in light nuclei}


\author{P.-G. Reinhard}
\affiliation{Institut f\"ur Theoretische Physik, Universit\"at Erlangen, D-91054 Erlangen, Germany}
\author{J. A. Maruhn}
\affiliation{Institut f\"ur Theoretische Physik, Goethe-Universit\"at, D-60438 Frankfurt am Main, Germany}
\author{A. S. Umar}
\affiliation{Department of Physics and Astronomy, Vanderbilt University, Nashville, Tennessee 37235, USA}
\author{V. E. Oberacker}
\affiliation{Department of Physics and Astronomy, Vanderbilt University, Nashville, Tennessee 37235, USA}

\date{\today}


\begin{abstract}

We investigate the presence of spatial localization in nuclei using a method that
maps the nucleon same-spin pair probability and is based on the density-matrix. 
The method is used to
study spatial localization of light nuclei within the Hartree-Fock approximation.
We show that the method provides an alternative tool for studying spatial localization in comparison
to the localization observed from maxima in the nuclear mass density.
\end{abstract}
\maketitle


\section{Introduction}

Clustering phenomena in light nuclei have always been an intriguing
aspect of nuclear structure physics. Theoretical understanding of why
and how conglomeration of nucleons to subunits within a nucleus
results in an increase in stability remains an actively investigated
question.  In particular, alpha clustering in light nuclei has a long
history~\cite{HT38,Mo56,Ik68,Bri68a} and suggests the existence of
configurations resembling the formation of \textit{nuclear
molecules}~\cite{IO01,IO06,WO06}.  It has also been suggested that
neutron rich isotopes of some light nuclei may give rise to new types
of cluster structures~\cite{II08,WO06}.  Most of the theoretical
analyses for the cluster structures have been performed with
the \textit{a priori} initialization in terms of clusters and
effective interactions, which are determined such as to reproduce the
binding energies and scattering phase shifts of these configurations.
On the other hand, nuclear structure calculations based on the
independent-particle approximation or density functionals also
manifest cluster-like substructures as marked concentration of
density in the visualization of the total nuclear mass density. For
example, Hartree-Fock (HF) calculations for light nuclei often show
such formations~\cite{Mar10a}, however since the HF single-particle
states are generally spread across the whole nucleus they are
delocalized, which makes the entanglement of these substructures in
terms of the single-particle orbitals very difficult.  Furthermore,
the identification of cluster and shell structures based only on the
mass density may be an oversimplification since it is missing
other aspects of the many-body system, for example the kinetic energy
density or density gradients, which may help to provide a more
detailed understanding of the underlying structure. Finally,
with the rising popularity of the density functional approach in
nuclear physics it may be desirable to have a new localization measure
that stems directly from the nuclear density-matrix, since all of the
information is contained in this quantity.

\section{The localization measure}

\subsection{Outline of formalism}
An alternative measure of localization 
had been developed in the context of
a mean-field description for electronic systems ~\cite{BE90}.  A fermionic
mean-field state is fully characterized by the one-body density-matrix
\begin{equation}
  \rho_{q\sigma\sigma'}(\mathbf{r},\mathbf{r'})
  =
  \sum_{\alpha\in q} 
  \phi_{\alpha}(\mathbf{r}\sigma)\phi_{\alpha}^{*}(\mathbf{r'}\sigma')
  \;.
\end{equation}
The probability of finding two nucleons with the same spin at spatial
locations $\mathbf{r}$ and $\mathbf{r'}$ (same-spin pair probability) for
isospin $q$ is given by
\begin{equation}
  P_{q\sigma}(\mathbf{r},\mathbf{r}')
  =
  \rho_{q\sigma}(\mathbf{r})\rho_{q\sigma}(\mathbf{r'})
  -
  \left|\rho_{q\sigma\sigma}(\mathbf{r},\mathbf{r'})\right|^2
  \;,
\end{equation}
where
$\rho_{q\sigma}(\mathbf{r})=\rho_{q\sigma\sigma}(\mathbf{r},\mathbf{r})$ is
the local density. The conditional probability for finding a nucleon at
$\mathbf{r'}$ when we know with certainty that another nucleon with the same
spin and isospin is at $\mathbf{r}$ is
\begin{equation}
  R_{q\sigma}(\mathbf{r},\mathbf{r}')
  =
  \rho_{q\sigma}(\mathbf{r'})
  -
  \frac{\left|\rho_{q\sigma\sigma}(\mathbf{r},\mathbf{r'})\right|^2}
       {\rho_{q\sigma}(\mathbf{r})}
  \;.
\end{equation}
Since we are interested in the localization aspects of this probability it is
sufficient to consider only the local short-range behavior of the conditional
probability, which one can obtain by performing a spherical averaging over a
shell of radius $\delta$ about the point $\mathbf{r}$ and then Taylor expanding
the resulting expression to get~\cite{BE90}
\begin{equation}
\label{taylor}
  R_{q\sigma}(\mathbf{r},\delta)
  \approx
  \frac{1}{3}\left(
   \tau_{q\sigma}
   -
  \frac{1}{4}
  \frac{[\mathbf{\nabla}\rho_{q\sigma}]^2}{\rho_{q\sigma}}
  -
  \frac{{\bf j}_{q\sigma}^2}{\rho_{q\sigma}}\right)\delta^2
  +
  \mathcal{O}(\delta^3)
  \;,
\end{equation}
where $\tau_{q\sigma}$ and ${\bf j}_{q\sigma}$ are the kinetic energy density
and current density given by
\begin{eqnarray*}
  \tau_{q\sigma}(\mathbf{r})
  &=&
  \sum_{\alpha\in q}
  \left|\mathbf{\nabla}\phi_{\alpha}(\mathbf{r}\sigma)\right|^2 
\\
  {\bf j}_{q\sigma}(\mathbf{r})
  &=&
  \sum_{\alpha\in q}
  \mathrm{Im}\left[
    \phi_{\alpha}^{*}(\mathbf{r}\sigma)
     \mathbf{\nabla}\phi_{\alpha}(\mathbf{r}\sigma)
  \right]
\\
  \mathbf{\nabla}\rho_{q\sigma}(\mathbf{r})
  &=&
  2\sum_{\alpha\in q}
  \mathrm{Re}\left[
    \phi_{\alpha}^{*}(\mathbf{r}\sigma)
     \mathbf{\nabla}\phi_{\alpha}(\mathbf{r}\sigma)
  \right]
  \;.
\end{eqnarray*}
The reason for writing $\mathbf{\nabla}\rho_{q\sigma}$ explicitly is
to emphasize that to have a smooth behavior of the quantities
calculated below it is essential to calculate all quantities directly
from the wavefunctions.  The expression shown in Eq.~(\ref{taylor})
suggests the definition of a localization measure
\begin{equation}
\label{dsig}
D_{q\sigma}(\mathbf{r})=\left(\tau_{q\sigma}-\frac{1}{4}\frac{[\mathbf{\nabla}\rho_{q\sigma}]^2}{\rho_{q\sigma}}-\frac{{\bf j}_{q\sigma}^2}{\rho_{q\sigma}}\right)\;,
\end{equation}
which is also valid for time-dependent Slater
determinants~\cite{BM05}.
It is important to remember that $D_{q\sigma}$ is the short-range limit of the
conditional like-spin \textit{pair} probability and may contain correlations
that are not evident in simple one-body observables, such as the mass
density.  The localization measure defined by Eq.~(\ref{dsig}) is a reverse
relation, e.g. the larger the probability of finding two like-spin particles
in vicinity of each other the smaller the value of $D$. For this reason it is
customary to define a reversed and normalized localization measure
\begin{eqnarray}
\label{eq:local}
  {\cal C}_{q\sigma}(\mathbf{r})
  &=&
  \left[
  1
  +
  \left(
  \frac{\tau_{q\sigma}\rho_{q\sigma}-\frac{1}{4}[\nabla\rho_{q\sigma}]^2-{\bf j}_{q\sigma}^2}
       {\rho_{q\sigma}\tau_{q\sigma}^\mathrm{TF}}
  \right)^2
  \right]^{-1}
\\
  \tau_{q\sigma}^\mathrm{TF}
  &=&
  \frac{3}{5}\left(6\pi^2\right)^{2/3}\rho_{q\sigma}^{5/3}
  \;,
\nonumber
\end{eqnarray}
where $\tau_{q\sigma}^\mathrm{TF}$ is the Thomas-Fermi kinetic energy
density.
The latter is used to provide a natural scale which then allows to
define a dimensionless measure.
The current density vanishes in the static case which we will consider
in the following.

\subsection{Limiting cases and interpretation}

This criterion (\ref{eq:local}) is known in electronic systems as
electron localization function (ELF) and it is used as one ingredient
to analyze the bond structure of molecules in the static \cite{BE90}
and dynamic domain \cite{BM05}. The information content of the
localization function is understood from considering limiting cases.

The extreme case of ideal metallic bonding is
realized for homogeneous matter where $\tau=\tau_{q\sigma}^\mathrm{TF}$.  This
yields $\mathcal{C}=\frac{1}{2}$, a value which thus signals a region with a
nearly homogeneous Fermi gas as it is typical for metal electrons, nuclear
matter, or neutron stars. 
The opposite regime are space regions where exactly one single-particle
wavefunction of type $q\sigma$ contributes.  This is
called \textit{localization} in molecular physics.  Such a situation
yields
$D_{q\sigma}(\mathbf{r})=0$,
since it is not possible to find another like-spin state in the vicinity,
and consequently $\mathcal{C}=1$, the value which
signals \textit{localization}. 
It should be noted that the localization function is invariant
under unitary transformations amongst the single-particle
wavefunctions in a Slater state~\cite{Sav05}.
In the nuclear case, it is the
$\alpha$ particle which is perfectly localized in this sense,
i.e. which has $\mathcal{C}=1$ everywhere for all states.
Well bound nuclei show
usually metallic bonding and predominantly have
$\mathcal{C}=\frac{1}{2}$. Light nuclei are often expected to contain
pronounced $\alpha$-particle sub-structures. 
Such a sub-structure means that in a certain region of space only
an $\alpha$ particle is found which in turn is
signaled by $\mathcal{C}=1$ in this region.  In fact, an $\alpha$
sub-structure is a correlation of four particles: $p\uparrow$,
$p\downarrow$, $n\uparrow$, and $n\downarrow$. Thus it is signaled
only if we find simultaneously for all four corresponding localization
functions ${\cal C}_{q\sigma}\approx 1$. In the following, we will
consider mainly $N=Z$ nuclei for which the four different particles
have very similar wavefunctions. In this case,  it suffices to
consider, pars pro toto, only one localization function.
Furthermore, it should be noted that a full identification of
$\alpha$-cluster sub-structures requires also to check the
correlations between the four nucleons gathering in a ``localized''
region  of the nucleus. The localization function is just the
first step to identify those regions, namely the minimum necessary
condition. 

\section{Results and discussion}

In our calculations, the static HF equations are solved on a Cartesian
three-dimensional mesh without any symmetry assumptions. The grid spacing was
$1$~fm with a box size of $(-15.5,+15.5)$~fm in each dimension.  The Skyrme
energy functional was employed with the parametrization SkI3~\cite{RF95}.  The
spatial derivatives are calculated using the fast Fourier transform and
periodic boundary conditions are employed, except for the Coulomb potential,
which is calculated with boundary conditions at infinity as described in
Ref.~\cite{EB79}. 

\subsection{Ground states of $N=Z$ nuclei}

\begin{figure}[!htb]
\includegraphics*[width=8.6cm]{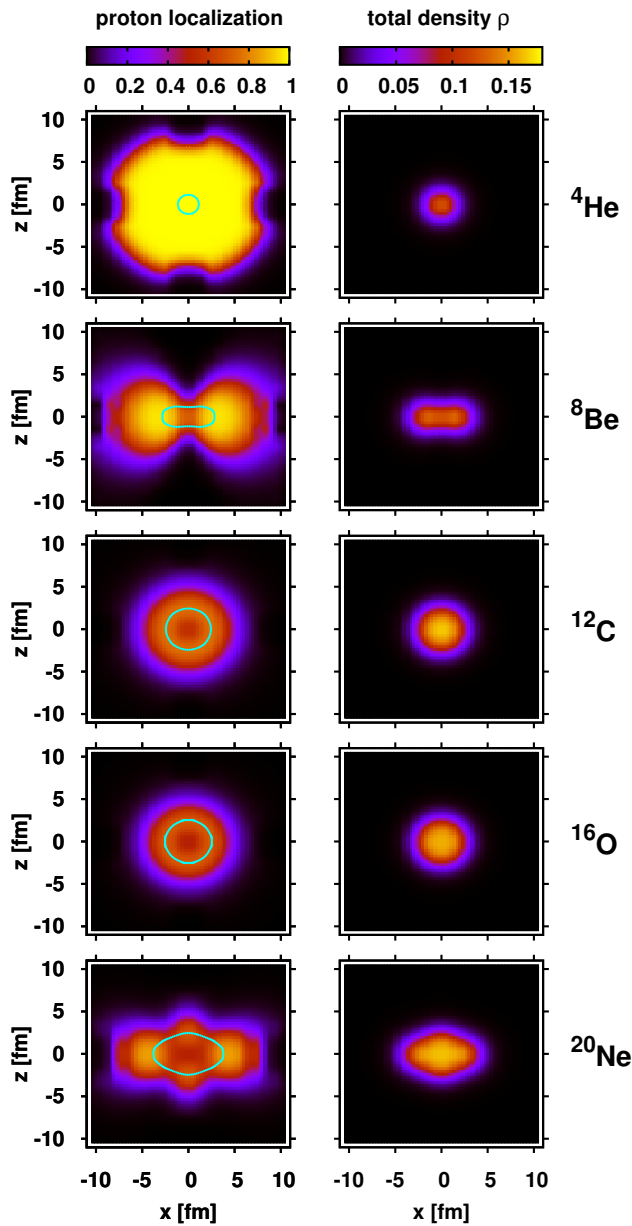}
\caption{\label{fig1}
(Color online)
Color map (gray scale) plots of
proton localization (left column) and total density in fm$^{-3}$ (right column),
for the $Z=N$ nuclei up to $^{20}$Ne.
The position of the density contour at half nuclear matter density
($\rho=0.08$ fm$^{-3}$) is indicated with color cyan (light gray)
in the maps of proton localization.
}
\end{figure}
\begin{figure}[!htb]
\includegraphics*[width=8.6cm]{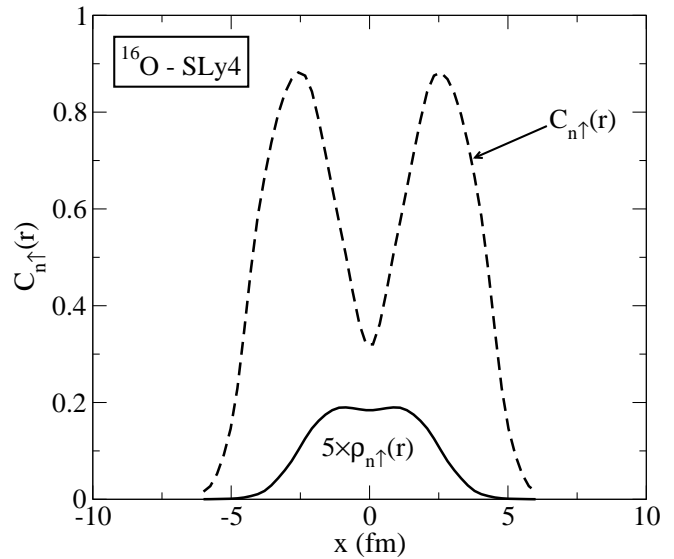}
\caption{\label{fig2}
(Color online)
Density profile and localization function for the $^{16}$O
nucleus.
}
\end{figure}

Fig.~\ref{fig1} shows an $x$-$z$-cut of the localization function
(\ref{eq:local}) for even-even $N=Z$ nuclei from $A=4$ to $A=20$.  The
left panel shows the proton localization criterion
$\mathcal{C}_{p\uparrow}$ complemented in the right panel by the
corresponding total density. As mentioned above, the states are
spin symmetric which yields identical localization plots for spin-up
and spin-down. Moreover, for light $N=Z$ nuclei proton and neutron
localizations are very similar due to the small Coulomb
interaction. (For neutron rich isotopes this is no longer true as we
will show below.)  The color (gray scale) coding is shown on top of
each column and remains the same throughout the column.  The position
of the density contour at half nuclear matter density ($\rho=0.08$
fm$^{-3}$) is indicated with color cyan in the maps of proton
localization.  One should keep in mind that the maxima and minima of
the total nuclear density need not be correlated with that of the
localization function, which is a topological quantity to describe
localization (see also Fig. \ref{fig2} and discussion thereof).
The top panel of Fig.~\ref{fig1} shows the calculations for the
$^{4}$He nucleus.  As we have described previously we see a perfect
localization with $\mathcal{C}=1$ in all relevant regions where
$\rho>10^{-4}$ fm$^{-3}$. Smaller densities lead to erroneous results
for $\mathcal{C}$ due to the very subtle cancellations required.
The strongly prolate $^{8}$Be shows
very distinct localization pattern with perfect localization in the left and
right halves of the contour plane and much smaller localization
in the contact region where the wavefunctions overlap.
As can be seen this is much more pronounced in comparison to the total
mass density plot. Here, it is probably reasonable to conclude that
 $^{8}$Be can be considered as an  $\alpha$-$\alpha$ molecule.
With this version of the Skyrme force the ground state of $^{12}$C
is oblate deformed as shown in the right pane of
Fig.~\ref{fig1}. One may be tempted to consider this as a planar
arrangement of three $\alpha$ particles. A slight indication of that may
be spotted in the localization plot. But it is not well
developed, the configuration is too compact, and shows preferably
metallic binding as we can see from the localization (left
column) which stays safely in a regime  $\mathcal{C}\approx 1/2$.
The strongly bound $^{16}$O nucleus mostly shows a localization value
of $\mathcal{C}=\frac{1}{2}$ throughout as one would have
expected. Its density is known to have a dip at the
center \cite{Fri82a}. This cannot be discriminated in the density
plot here but can be observed as a region of lower localization in the
localization map plot.
To examine this further we have repeated the same calculation for $^{16}$O
using the SLy4 interaction~\cite{CB98}. In Fig.~\ref{fig2} we show a
cut through the profile of the density and similarly through 
the localization function.
We observe that the central dip in the total density is barely
visible. The localization function, however, shows a very pronounced dip
indicating a strong and irreducible overlap of all wavefunctions in
this center region. Note, furthermore, that the maxima of mass
density and localization do not coincide. The localization
has a preference towards the surface where the lower density
enhances the chance of finding one prevailing wavefunction.

Finally, the last panel of Fig.~\ref{fig1} shows results
for the strongly prolate $^{20}$Ne nucleus. The localization map
shows two regions of high localization at the outer ends and a
ring of somewhat enhanced localization at the center around the
elongation axis. One can interpret this as a quasi-molecular
$\alpha$-$^{12}$C-$\alpha$ configuration. The $\alpha$ substructures
on both sides are almost as well developed as in $^8$Be.
We have also  computed the further series of $N=Z$ nuclei,
$^{24}$Mg, $^{28}$Si, $^{32}$S, $^{36}$Ar, and $^{40}$Ca. These nuclei
are increasingly compact and all show basically metallic binding
similar to $^{12}$C and $^{16}$O shown here.

\subsection{Strongly deformed configurations of light $N=Z$ nuclei}

\begin{figure}[!htb]
\includegraphics*[width=8.6cm]{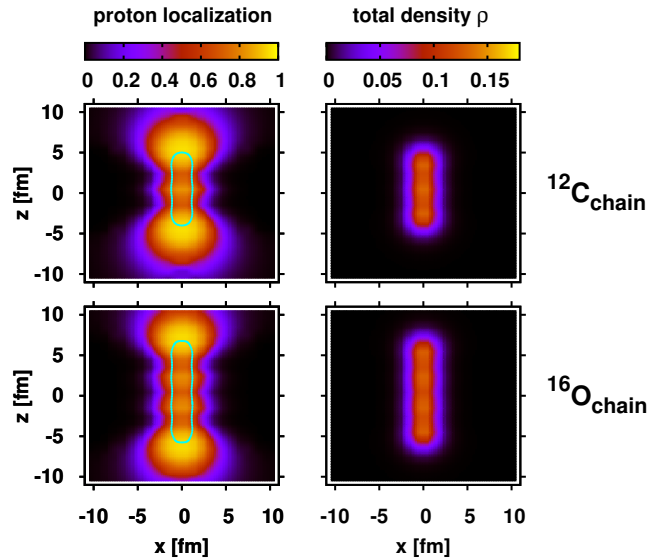}
\caption{\label{fig3}
(Color online)
As figure \ref{fig1}, but for chain-like isomers of
$^{12}$C and $^{16}$O.
}
\end{figure}
Very light $N=Z$ nuclei are likely to display isomeric, or at least
transiently stable, configurations which are very elongated and
resemble chains of $alpha$ particles \cite{Ho94}. For somewhat
heavier $N=Z$ one finds often shape coexistence with strongly prolate
deformed nuclear configurations \cite{Re99}. Such less bound and
spatially more extended configurations are more likely to allow for
$\alpha$ sub-structures.
We thus have also considered such isomeric configurations for a number of
light $N=Z$ nuclei. 
These configurations were found numerically by starting the 
static iteration from a sufficiently prolate configuration
such that the iteration converged to 
the elongated isomeric state. Chain configurations were found
immediately for $^{12}$C and $^{16}$O while the heavier systems
preferred to maintain a compact core between the $\alpha$ satellites.
It  is to be noted that these configurations are stable minima in a
mean field calculation. They may hybridize with the ground state
in correlated calculations. Still such configurations may show up as
transient configurations in nuclear reactions  \cite{Ho94}.

Fig.~\ref{fig3} shows the total density and localization plots
for the linear-chain states of $^{12}$C and $^{16}$O nuclei.  For both
the density suggests an $\alpha$-chain structure which
is, indeed, corroborated by the localization that also shows three or
four clearly separated maxima, $\mathcal{C}\approx 1$. The region of
high localization is very large at both ends, but much smaller for
the maxima in the interior due to larger wavefunction overlap.
One interesting point about the $^{12}$C linear-chain configuration
localization plot is that in studying the dynamical formation of this chain
state, as it was done in Ref.~\cite{UM10}, we have observed that the
dynamical vibrations of the mass density resembled the localization plot
with only the equilibrium shape having the triple-$\alpha$ structure.
This is consistent with the kinetic interpretation of the localization
function, suggesting that kinetic energies of the same-spin pairs peak mostly
around the ends of the linear-chain.

\begin{figure}[!htb]
\includegraphics*[width=8.6cm]{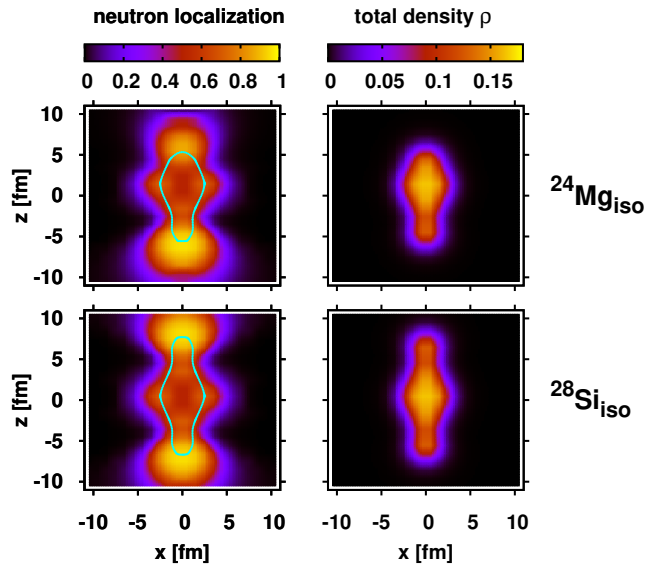}
\caption{\label{fig4}
(Color online)
As figure \ref{fig1}, but for stretched isomers of $^{24}$Mg and $^{28}$Si.
}
\end{figure}
Fig.~\ref{fig4} shows strongly prolate (not yet chain-like isomers which lie
higher in energy) isomers of
$^{24}$Mg and $^{28}$Si. Unlike the compact ground-state configurations
these isomers indicate interesting molecular substructures. One may
interpret $^{24}$Mg as a $\alpha$-$^{12}$C-$\alpha$-$\alpha$ molecule
and $^{28}$Si as  $\alpha$-$\alpha$-$^{12}$C-$\alpha$-$\alpha$.
Again, the outermost $\alpha$'s are best developed with large regions
of high localization. The inner $\alpha$'s have already degraded 
localization due to neighboring wavefunctions from both sides.

\subsection{An example for $N>Z$: The $^{20}$C chain}

\begin{figure}[!htb]
\includegraphics*[width=8.6cm]{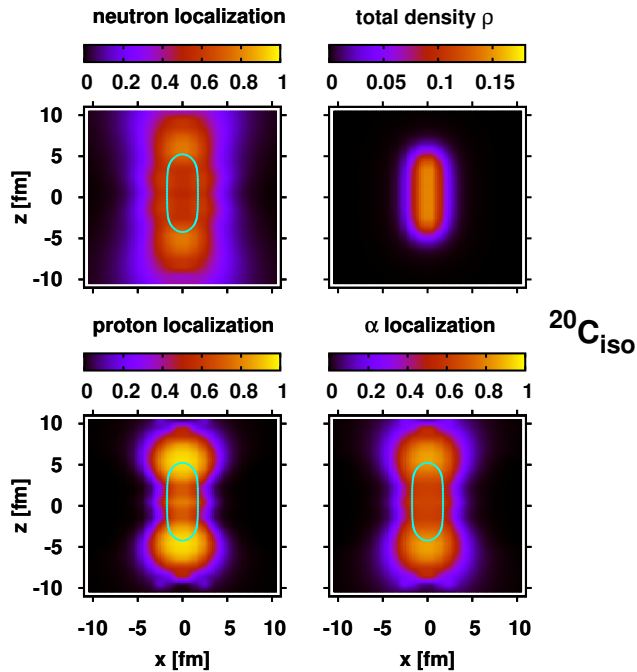}
\caption{\label{fig5}
Color map plots of localizations and density 
for the linear-chain configuration of $^{20}$C.
Lower left: proton localization.
Upper left: neutron localization.
Lower right: $\alpha$ localization 
($\equiv\sqrt{\mathcal{C}_{p\uparrow}\mathcal{C}_{n\uparrow}}$).
Upper right: total density (in fm$^{-3}$).
The position of the density contour at half nuclear matter density
($\rho=0.08$ fm$^{-3}$) is indicated with color cyan (light gray) in the
maps of localization.
}
\end{figure}
Recently, much interest has been devoted to the study of cluster
configurations for neutron-rich isotopes of light
nuclei~\cite{IO01,IO06,ML10}.  In particular the linear-chain
configurations of C isotopes and their stability against bending modes
has been of interest.  
For nuclei with $N>Z$ where proton and neutron
wavefunctions are naturally different the search for $\alpha$
sub-structure requires a simultaneous analysis of proton and neutron
localization. To that end we consider also as $\alpha$ localization
the combination
$\sqrt{\mathcal{C}_{p\uparrow}\mathcal{C}_{n\uparrow}}$.  The spin-up
and spin-down wavefunctions are still degenerate such that it suffices
to consider one of the spins.
In Fig.~\ref{fig5} we show proton,
neutron, and $\alpha$ localization plots for the linear-chain isomer
of the $^{20}$C nucleus. 
As expected,
due to the neutron excess of $^{20}$C the localization plots for
neutrons and protons look considerably different.  The protons show
more distinct regions with high localization value in comparison to
the neutron case, where the wavefunctions have more overlap due
to the large number of neutrons.  
The $\alpha$ localization is the obvious average of the
two left panels. In spite of the neutron cloud from the excess
neutrons, there appears still some faint $\alpha$ sub-structure
at the edges of the chain.
It is also interesting to observe
that the total mass density does not show any pronounced
features due to the smoothing effect of the surplus neutrons while the
localization plots still reveal noteworthy structures.

\medskip

\section{Conclusion}

In summary, we have applied a localization measure
which was developed originally for analyzing bonding structures in
molecules to a study of $alpha$ sub-structures in light nuclei.
The localization function
is obtained directly from the density-matrix, in the mean-field
approximation. It depends on kinetic-energy
density and current density, in addition to the mass density.
It can be easily implemented
for density functional theory calculations of nuclear structure.
One of the fundamental reasons why the new localization measure is such
an excellent predictor of correlation and localization is due to
the fact that it incorporates the kinetic energy of the relative
motion of spin-parallel nucleons at a particular point in space
in addition to the mass density for the system~\cite{Do91}.
In most cases this localization function shows more
detailed localization or clustering features in comparison to the total
mass density.
Results for $N=Z$ nuclei up to $^{40}$Ca show that pronounced
localization, associated with $\alpha$-particle substructures, appear
only for the strongly prolate ground states of $^{8}$Be,
$^{20}$Ne, and of course trivially for $^4$He. All other nuclei are
more compact and show metallic binding. However, stretched isomers
of light nuclei often show convincing $\alpha$ structures, 
particularly well developed
for the $\alpha$ chains of $^{12}$C and $^{16}$O, but also for the
prolate $^{24}$Mg and $^{28}$Si isomers.
In the future we also plan to study the new localization function in
time-dependent HF calculations of systems related to nuclear molecular
configurations.

This work has been supported by the U.S. Department of Energy under grant No.
DE-FG02-96ER40963 with Vanderbilt University, and by the German BMBF
under contract Nos. 06FY9086 and 06ER142D.

\end{document}